\documentclass[useAMS,usenatbib]{mn2e}
\usepackage{epsfig}
\usepackage{epstopdf}
\usepackage{times}
\usepackage[fleqn]{amsmath}
\usepackage{amssymb}
\usepackage{amsbsy}

\title[Extracting the global 21-cm cosmic dawn signal]{An MCMC
approach to extracting the global 21-cm signal during the cosmic dawn
from sky-averaged radio observations}

\author[G.\ J.\ A.\ Harker, J.\ R.\ Pritchard, J.\ O.\
    Burns and J.\ D.\ Bowman]{Geraint J.\ A.\ Harker,$^{1,3}$\thanks{Email:
    geraint.harker@colorado.edu} Jonathan R.\ Pritchard,$^{2}$ Jack
    O.\ Burns$^{1,3}$ and \newauthor Judd D.\ Bowman$^{4}$\\ $^{1}$Center for Astrophysics and Space
    Astronomy, Department of Astrophysics and Planetary Sciences,
    University of Colorado at Boulder, CO 80309, USA\\
    $^{2}$Harvard--Smithsonian Center for Astrophysics, 60 Garden
    Street, Cambridge, MA 02138, USA\\
    $^{3}$NASA Lunar Science Institute, NASA Ames Research Center, Moffett Field, CA 94035, USA\\
    $^{4}$Arizona State University, School of Earth and Space Exploration, Tempe, AZ 85287, USA
}

\begin{document}

\date{\today}

\maketitle

\begin{abstract}
Efforts are being made to observe the 21-cm signal from the `cosmic
dawn' using sky-averaged observations with individual radio dipoles.
In this paper, we develop a model of the observations accounting for
the 21-cm signal, foregrounds, and several major instrumental effects.
Given this model, we apply Markov Chain Monte Carlo techniques to
demonstrate the ability of these instruments to separate the 21-cm
signal from foregrounds and quantify their ability to constrain
properties of the first galaxies. For concreteness, we investigate
observations between 40 and 120 MHz with the proposed {\it DARE}
mission in lunar orbit, showing its potential for science return.
\end{abstract}

\begin{keywords}
cosmology: theory -- diffuse radiation -- methods: statistical --
radio lines: general
\end{keywords}

\section{Introduction}\label{sec:intro}

One of the remaining frontiers of modern cosmology is the end of the
`dark ages' and the `cosmic dawn'.  This is the period ranging from
roughly 100 million years ($z\sim30$) to a billion years ($z\sim6$)
after the Big Bang, when the first stars and galaxies formed, lighting
up the Universe.  This period lies at the edge of current
observational techniques and is of considerable theoretical interest.
The next decade is expected to see significant improvements in
observations as telescopes such as the {\em James Webb Space
Telescope} ({\it{}JWST}) and the Atacama Large Millimeter Array (ALMA)
go online.  These instruments will provide considerable information
about galaxy formation at $z\lesssim 10$--$15$, but even these large
telescopes will be hard pressed to probe the very beginning of the
cosmic dawn.

Measurements of 21-cm emission and absorption from intergalactic
hydrogen at high redshift promise to increase greatly our knowledge of
the Universe at redshifts $z\gtrsim 6$ \citep*{madau1997}. Experiments
under way at present are concentrating on frequencies $\nu\gtrsim 100\
\mathrm{MHz}$ ($z\lesssim 13.2$), and are hoping to capture the
transition from an almost completely neutral Universe to an almost
completely ionized one (the epoch of reionization, or EoR). Future
observations at yet lower frequencies (higher redshifts) may probe the
epoch when the first sources formed -- `cosmic dawn' -- and even the
preceding `dark ages'.

There are two main approaches to making these measurements: using a
large interferometric array to produce statistics (e.g.\ power
spectra), and perhaps even images, of the 21-cm brightness
temperature; or using a single antenna to measure the mean brightness
temperature as a function of frequency and redshift \citep{SHA99}. In
either case, the bright foregrounds at low frequencies present one of
the most significant challenges to extracting the 21-cm signal. The
difficulty is alleviated somewhat in the former approach since an
interferometer is sensitive only to fluctuations in the foregrounds,
which are small compared to the mean on the scales of interest, but
they still exceed the 21-cm fluctuations in intensity by several
orders of magnitude. Interferometric measurements have other benefits
too. For example, the spectrum of fluctuations carries more
information than the mean signal alone, and interferometers may make
it easier to identify and excise man-made radio-frequency interference
(RFI).  An interferometer cannot measure the mean signal, however.
Moreover, global signal experiments designed to measure the mean
brightness temperature may be much simpler and cheaper than large
arrays, and are not troubled to the same extent by distortions caused
by the Earth's ionosphere.

The large sky temperature at these frequencies also means that the sky
makes the dominant contribution to the system temperature, and hence
to the sensitivity of the observation for a given bandwidth and
integration time. The brightness temperature, $T_\mathrm{B}$, of the
diffuse foregrounds depends on the observing frequency, $\nu$, as
$T_\mathrm{B}\sim\nu^{-2.5}$ \citep{ROG08}, so interferometric
measurements during the cosmic dawn at less than 100 MHz require very
long integration times or arrays with a very large collecting
area. Because ionospheric effects also become more serious at low
frequencies, it has been suggested that the far side of the Moon,
which is also free (as yet) from RFI, would be the best and perhaps
the only site for an array to probe the cosmic dawn and dark ages
\citep[e.g.][]{burns1988,burns2009,jester2009}.  Building and
operating such an array of the requisite size would be quite a
formidable undertaking, so global signal experiments provide the best
hope for probing the 21-cm signal at $z\gtrsim 15$ in the near future.

The EDGES experiment \citep{BOW10}, operating at 100--200 MHz, has
pioneered global 21-cm measurements, recently placing limits on how
rapidly the global 21-cm signal may vary with frequency, and thereby
putting a lower limit on the duration of the reionization epoch.  Even
from its superb radio-quiet site in Western Australia, however, it
encountered RFI from sources such as telecommunications satellites and
radio and television transmitters. The signals from these may reach
EDGES quite directly, or arrive via e.g.\ tropospheric scattering or
reflections from aircraft and meteor trails.  This requires a large
fraction of the data to be discarded, which would be more damaging at
low frequencies where longer integrations are required, and it imposes
stringent demands on the dynamic range of the receiver.

A dipole antenna in orbit around the Moon could avoid these problems,
since it would be free of RFI when shielded from the Earth over the
lunar farside. In addition, an antenna in space experiences a simpler
and more stable environment than one on the Earth's surface, which may
allow for more straightforward calibration. The use of lunar orbit
does not require anything to be landed on the Moon's surface, unlike
for a farside array. Such a mission concept has been developed, called
the {\it Dark Ages Radio Explorer} \citep[{\it
DARE};][]{burns2011}\footnote{http://lunar.colorado.edu/dare/}. In
this paper we therefore explore the constraints that a mission such as
{\it DARE} could place on a model of the 21-cm brightness temperature
between the end of the cosmic dark ages and the start of the epoch of
reionization. We aim to include all the most important contributions
to the low-frequency radio spectra measured by a dipole in lunar
orbit: the redshifted 21-cm signal itself; spatially varying diffuse
foregrounds based on an empirical model of the low-frequency radio
sky; the Sun; the thermal emission of the Moon; the reflection of
emission from other sources by the Moon; the response of the
instrument, which is based on an electromagnetic model of an antenna
design proposed for the {\it DARE} mission; and the thermal noise for
a realistic mission duration. Parameters describing all these
components must be fit simultaneously from the data since, for
example, it may be that the properties of the instrument cannot be
computed or measured on the ground with sufficient accuracy to allow
recovery of the 21-cm signal in the presence of the very bright
foregrounds.

We therefore extend the work of \citet{PRI10}, who used Fisher matrix
and Monte Carlo methods to predict the accuracy with which models of
the 21-cm signal could be constrained by a single antenna, but who
considered the simpler case of an experiment which measured a single,
deep spectrum (i.e.\ they did not consider the variation of
foregrounds over the sky), and where the only contributions to the
measured spectrum were the redshifted 21-cm signal, diffuse
foregrounds and noise.

The techniques that we develop, and the basic form of our model for
the 21-cm global signal, are quite generic and may be applied to
future experiments both on the ground and in space.  For concreteness,
we focus here on the proposed {\it DARE} mission, but a similar
methodology could be applied to EDGES and other ground based
experiments.  We plan to investigate this in the near future.

We start by outlining the relevant features of our reference
experiment, a proposed mission to measure the 21-cm global signal from
lunar orbit, in Section~\ref{sec:mission}. Then, in
Section~\ref{sec:sims} we describe all the different effects which are
included in our simulations of data from such a mission, including the
parametrizations we use. We also discuss some other contributions,
such as impacts of exospheric dust on the antenna and radio
recombination lines, and justify neglecting them in this analysis.

Constraints on the model parameters are derived using a Markov Chain
Monte Carlo (MCMC) method. In Section~\ref{sec:fitting} we introduce
our implementation of this technique and, as an example, show how well
the parameters are recovered by a perfect instrument, which is
sensitive across the whole frequency band and whose properties are
known exactly. In Section~\ref{sec:disc} we consider a more realistic
instrument with an imperfectly known response, and look at the impact
of the various processes we model on the quality of our
constraints. Finally, we offer some conclusions in
Section~\ref{sec:conc}.

\section{Reference experiment}\label{sec:mission}

We base our simulations on the proposed {\it DARE} mission, a fuller
description of which will be given by \citet{burns2011}, and which
acts as our reference experiment. {\it DARE} is designed to carry a
low-frequency radio antenna in a circular, equatorial orbit 200~km
above the surface of the Moon. Data would only be taken during the
part of the orbit when the Moon blocks RFI from the
Earth. Approximately 30~min out of each 127~min orbit is spent out of
direct line of sight of the Earth and outside the diffraction zone of
terrestrial RFI around the lunar limb. A conservative estimate for the
total amount of useful integration time for a mission duration of
three years is 3000~h.

The primary data product will be a series of spectra at 40--120~MHz
with an integration time of 1~s, and with a spectral resolution of
around 10~kHz. The analysis in this paper assumes that these spectra
have been combined into spectra with a resolution of 2~MHz in a number
of discrete sky regions. This could be done either by taking discrete
pointings in different directions, and integrating for a long time in
each direction, or by scanning the pointing direction across the sky
and performing a map-making procedure to combine the
individual spectra together. In this paper we simulate only the final,
integrated spectra, not the individual high-resolution spectra or the
process of combining them.

\begin{figure}
  \begin{center}
    \leavevmode
    \psfig{file=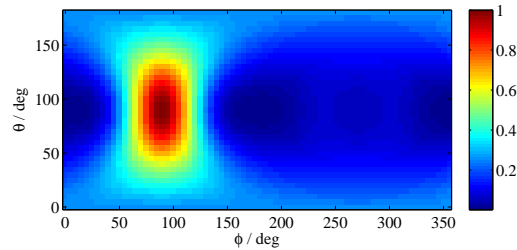,width=8cm}
    \caption{The simulated power pattern of one of the pair of dipoles
    which constitutes the {\it DARE} antenna at 75 MHz, plotted as a
    function of angle on the sky. The response is normalized to unity
    at its maximum, and the coordinate system is chosen such that the
    antenna points towards the positive y-axis
    ($\theta=\phi=90^\circ$). The pattern appears stretched in the
    $\theta$ direction, but the combined pattern with the other dipole
    oriented at right-angles to the first is more
    symmetric.}\label{fig:darebeam}
  \end{center}
\end{figure}

The antenna consists of a pair of tapered, biconical, electrically
short dipoles, designed by R. Bradley of NRAO to satisfy the
requirements of the {\it DARE} mission \citep{burns2011}. To increase
the directivity, a form of ground plane is provided by radials
extending out from the main body of the spacecraft. The design
provides a beam with a single primary lobe with a half-power beam area
of around 1--2~sr, depending on frequency (the full width at half
maximum of the power pattern at 75~MHz is $57\degr$), so that at any
one time the antenna is sensitive to radiation from a large fraction
of the sky. Despite the radials, the antenna has some sensitivity to
radiation from behind the spacecraft (a backlobe, diminished by
9--15~dB). The simulated power response of one of the dipoles as a
function of angle at 75~MHz is shown in Fig.~\ref{fig:darebeam}. The
model used to predict this pattern incorporates the design of the
antennas themselves and their support structures, the radials which
form the `ground screen', and the spacecraft structure itself.

The antenna and the receiver are designed to produce a smooth
frequency response. The primary method by which the foregrounds are
distinguished from the 21-cm signal is through the spectral smoothness
of the foregrounds, so it is essential that the receiving system does
not compromise this smoothness. The frequency response of the system
has been modelled, and is discussed further in
Section~\ref{subsec:inst}.

\section{Simulations of mission data}\label{sec:sims}

\begin{figure}
  \begin{center}
    \leavevmode
    \psfig{file=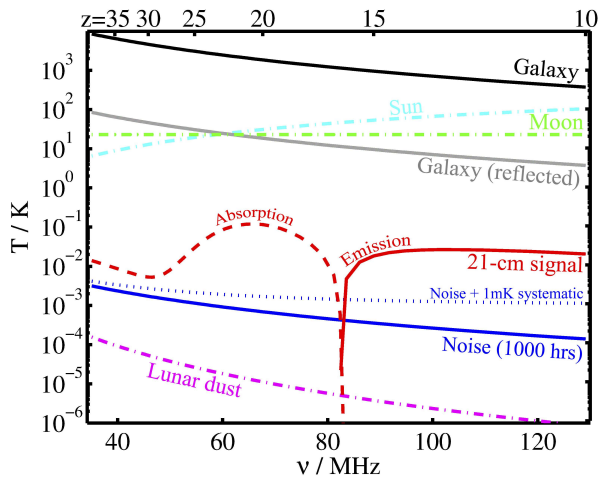,width=8cm}
    \caption{A comparison of the intensity of the 21-cm signal with
    that of various foregrounds and the thermal noise, as a function
    of frequency (bottom axis) and the corresponding redshift of the
    21-cm line (top axis). From the top (as they appear on the
    right-hand side of the plot), the different lines show spectra of:
    1) The diffuse foregrounds, from a region of sky away from the
    Galactic centre (solid black line); 2) The quiet Sun (dot-dashed
    cyan line); 3) The Moon, attenuated by being seen only through the
    backlobe of our simulated antenna (dot-dashed green line); 4) The
    diffuse foregrounds reflected by the Moon and entering the
    backlobe of the antenna (solid grey line); 5) The 21-cm signal
    (solid red line in emission, dashed red line in absorption); 6) \&
    7) Thermal noise after 1000 hours (solid blue line), and this
    noise with the addition of a 1mK systematic residual (dotted blue
    line); 8) Radio emission caused by the impact of dust particles
    from the lunar exosphere on the spacecraft and antenna (dot-dashed
    magenta line).}\label{fig:compare}
  \end{center}
\end{figure}

An overview of some of the different contributions to a spectrum
measured by low-frequency radio antenna in lunar orbit is given in
Fig.~\ref{fig:compare}.  Even when the antenna is oriented such that
it is sensitive mainly to an area of sky away from the Galactic
centre, the diffuse foregrounds (which consist mainly of synchrotron
radiation from our own Galaxy, with some contribution from free-free
emission and extragalactic sources; see e.g.\ \citealt{SHA99}) are
between four and six orders of magnitude brighter than the 21-cm
signal. Indeed, there are several other contributions which dominate
the 21-cm signal. In this section we describe our models for all these
contributions, and how they are combined into a simulation of the data
returned by a lunar-orbiting dipole experiment.

\subsection{The 21-cm signal}

The physics behind the properties of redshifted 21-cm emission and
absorption was reviewed by \citet*{FOB06}, and the evolution of the
21-cm signal with redshift (or cosmic time) was studied in more detail
by \citet{madau1997,ciardi2003,furlanetto2006,PRI08}. Of most interest
here is the redshift evolution of the sky-averaged (`global') signal.
More precisely, we look at the brightness temperature difference,
$\delta T_\mathrm{b}$, between 21-cm signal and the CMB at the
emission or absorption redshift, where $\delta T_\mathrm{b}<0$
indicates absorption against the CMB and $\delta T_\mathrm{b}>0$
indicates emission. This is given by
\begin{equation}
\begin{split}
\delta T_\mathrm{b} &= 27 x_{\mathrm{HI}}\left(\frac{T_\mathrm{S}-T_\gamma}{T_\mathrm{S}}\right)\left(\frac{1+z}{10}\right)^\frac{1}{2}\\
& \quad \times(1+\delta_\mathrm{b})\left[\frac{\partial_r v_r}{(1+z)H(z)}\right]^{-1}\,\mathrm{mK}\ ,
\end{split}
\label{eqn:deltatb}
\end{equation}
where $x_\mathrm{HI}$ is the hydrogen neutral fraction,
$\delta_\mathrm{b}$ is the overdensity in baryons, $T_\mathrm{S}$ is
the 21-cm spin temperature, $T_\gamma$ is the CMB temperature, $H(z)$
is the Hubble parameter, and the last term describes the effect of
peculiar velocities with $\partial_r v_r$ being the derivative of the
velocities along the line of sight. Because in this paper we consider
the sky-averaged signal, we will neglect its spatial fluctuations, so
that neither $\delta_\mathrm{b}$ nor the peculiar velocities will be
relevant and we are interested only in the spatial average of
$x_\mathrm{HI}$ and $T_\mathrm{S}$ in each redshift slice.

If the cosmological parameters are known, and absent a significant
heating effect from primordial magnetic fields \citep*{SCH09}, this
signal depends on the properties of the radiation which various
sources emit into the intergalactic medium, through its effects on
$x_\mathrm{HI}$ and $T_\mathrm{S}$. In principle, these sources can
include the decay or annihilation of dark matter particles
\citep*{FUR06a} or e.g.\ Hawking radiation from primordial black holes
\citep{MAC08}. We can be more confident, however, of there being a
significant contribution from stars and from accretion onto compact
objects such as black holes. UV radiation, coming primarily from
stars, couples the spin temperature of the 21-cm transition to the
kinetic temperature of the gas through the Wouthuysen-Field effect
\citep[`Ly$\alpha$ pumping';][]{WOU52,FIE58,FIE59}, while X-ray
radiation from black holes heats the gas
\citep{madau1997,mirabel2011}. It is likely that sufficient Ly$\alpha$
radiation is produced to couple the spin temperature to the kinetic
temperature well before sufficient X-rays are produced to heat the gas
above the CMB temperature and hence to put the 21-cm line into
emission \citep*{PRI08,ciardi2010}. The properties of early sources of
X-ray and Lyman alpha photons are highly uncertain, as is the star
formation history \citep{robertson2010}, making observations of the 21
cm global signal very valuable in learning about early galaxy
formation.

\begin{figure}
\begin{center}
\includegraphics[scale=0.4]{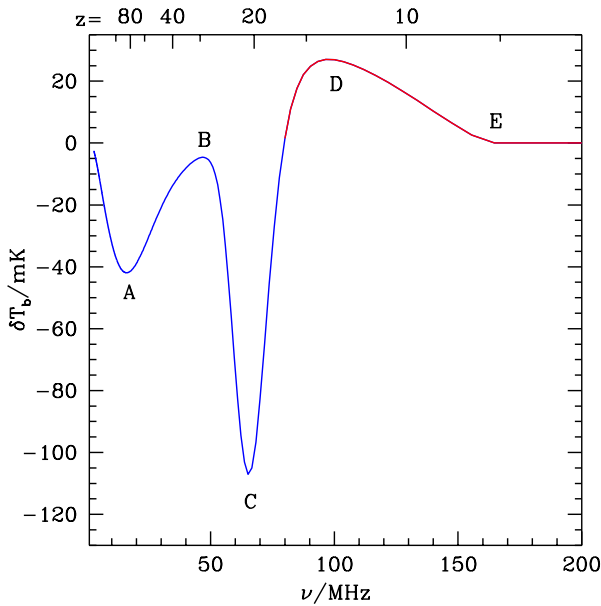}
\caption{Details of the 21-cm global signal as a function of
frequency, relative to the CMB, for our fiducial signal model.
The solid line shows the 21 cm global signal as it transitions from
absorption (blue) to emission (red). The different turning points are
labelled (see text for details).} \label{fig:nuhistory}
\end{center}
\end{figure}

The effect of varying the efficiency with which Ly$\alpha$ and X-rays
are produced and find their way into the IGM was studied by
\citet{PRI10}. The same paper proposed a useful parametrization of the
time-evolution of the global 21-cm signal, which we adopt here. The
signal is described by five turning points where the frequency
derivative of the signal is equal to zero (so each corresponds to a
local extremum of the signal). We label the turning points A--E (see
Figure \ref{fig:nuhistory}), in order from the highest to the lowest
redshift, and the physical interpretation of each is as follows:
\begin{description}
\item[A] -- a minimum during the dark ages where collisional coupling
  of the 21-cm spin temperature to the gas kinetic temperature begins
  to become ineffective;
\item[B] -- a maximum at the transition from the dark ages to the
  regime where Ly$\alpha$ pumping by UV from the first stars begins to
  become effective;
\item[C] -- a minimum as X-ray heating (caused by the first accreting
black holes) starts to become effective, raising the mean temperature;
\item[D] -- a maximum where the heating has saturated, before the
  signal begins to decrease because of cosmic expansion and
  reionization, i.e.\ the beginning of the EoR;
\item[E] -- the endpoint of reionization, after which the signal is (very
  close to) zero.
\end{description}
In this paper, the six parameters corresponding to the frequency and
$\delta T_\mathrm{B}$ of turning points B, C and D are varied, while
the the positions of A and E are fixed at $(16.1\ \mathrm{MHz},-42\
\mathrm{mK})$ and $(180\ \mathrm{MHz},0\ \mathrm{mK})$ respectively.
In the absence of exotic processes, turning point A depends only upon
fundamental cosmological parameters and known physics and so its
position is essentially known. Turning point E relates to the details
of the reionization history and, while its position is highly unknown,
here we focus on the first galaxies. The ability of global experiments
to constrain reionization has been considered in some detail by
\citet{PRI10} and \citet{MOR11}.

The 21-cm signal is modelled as a cubic spline interpolating these
points and having zero derivative at the position of the turning
points. We will consider constraints on these parameters to be the
primary result of an experiment to measure the global 21-cm signal at
these redshifts. Clearly, many other parametrizations are reasonable;
for example, we could attempt to constrain directly the input
parameters of a physical model for the global signal, such as the
spectral shape of early stars or the fraction of Ly$\alpha$ which
escapes early galaxies. We have chosen the `turning point'
parametrization because it is not as model-dependent, and because the
21-cm signal for a given set of parameters is very quick to compute,
which is desirable for our Monte Carlo analysis. Turning points B, C
and D, at around 45 MHz, 65 MHz and 100 MHz respectively, are visible
in the spectrum shown in Fig.~\ref{fig:path_alltight}, and are shown
in the context of a larger frequency range encompassing turning points
A and E in Fig.~\ref{fig:nuhistory}.

\subsection{Diffuse foregrounds}

Perhaps the most important foreground for a global 21-cm experiment,
in that it dominates in intensity and is present at some level for all
pointing directions, is the diffuse emission coming from our Galaxy
and external galaxies. Though the extragalactic foreground may be
considered to come from discrete sources, we treat it as part of
the diffuse foreground since the solid angle of the beam of our
proposed experiment is so large (around 1--2~sr) that it averages
together a great number of sources in any one pointing. The Galactic
contribution is larger than the extragalactic contribution, and
consists largely of synchrotron radiation, with a small contribution
from free-free \citep[e.g.][]{SHA99}.

Our model for the spatial variation of the diffuse foregrounds is the
global sky model (GSM) of \citet{DOC08}. The foreground temperature
measured when the spacecraft is pointing in a given direction is
obtained by convolving the GSM with the instrumental beam. We assume
that we can observe eight approximately independent sky areas (since
our simulated beam covers around one eighth of the sky, depending on
frequency), and that this is done by pointing in eight different
directions for equal amounts of time, with each direction being a
vertex of a spherical cube. This gives us eight foreground
spectra. These are modelled using a similar functional form to that
used by \citet{PRI10}, i.e.:
\begin{equation}
\begin{split}
\log T^i_\mathrm{FG}&=\log T^i_0 + a^i_1\log (\nu/\nu_0) +
a^i_2[\log(\nu/\nu_0)]^2 \\
& \quad + a^i_3[\log(\nu/\nu_0)]^3\ ,
\end{split}
\label{eqn:FGmodel}
\end{equation}
with $\nu_0=80\ \mathrm{MHz}$ being an arbitrary reference frequency
which we choose to lie in the middle of our band, and $i=1,\ldots ,8$
labels the different sky areas. The parameters
$\{T^i_0,a^i_1,a^i_2,a^i_3\}$ for $i=1,\ldots ,8$ constitute the 32
parameters of our diffuse foreground model. The `true' or input
parameters are obtained from fits to the GSM spectrum for each region.

This approach somewhat simplifies the problem since we have ignored
covariance between the different observed patches, treating them as
independent.  Instead, a realistic experiment would likely return a
sky map containing a larger number of correlated pixels with
approximately the same amount of information as our eight independent
pixels.  We leave a detailed study of map making and the handling of
correlated pixels to future work. However, it seems likely that the
overall effect of dealing properly with small correlations between the
patches would be to increase the error bars on the final constraints
slightly, since the foreground subtraction algorithm would have less
information to work with. In an extreme case with a single all-sky
integrated spectrum, the degeneracies between foreground, signal and
instrument parameters would clearly be severe. This contrasts with the
case for interferometric experiments, when knowledge of the
correlation properties of the foregrounds may help somewhat with
foreground subtraction since they have strong spatial correlations on
scales larger than the pixel size in an interferometric map
\citep{liu2011prd}. Correlations between pixels then work mainly to
increase the effective signal to noise ratio of the measurement of the
foreground in clusters of correlated pixels.

By restricting ourselves to a third order polynomial in $\log\nu$ in
each pixel, we are focusing on a relatively optimistic case.
\citet{PRI10} showed that this was the minimal number of parameters
needed to characterize the foreground model of \citet{DOC08}, but the
situation could be worse since this model is based on quite limited
observational data; \citet{petrovic2011} have, however, given
theoretical reasons to expect that the Galactic foregrounds should be
very smooth. Moreover, a recent study by \citet{liu2011} found that
four effective parameters was sufficient to fit a foreground model
with various components, with a total number of physical parameters
several times larger. The studies of \citet{PRI10} showed that
increasing the order of the polynomial required to fit the foregrounds
significantly worsened the constraining power of global 21 cm
experiments.  It would be straightforward to similarly explore the
effects of foregrounds with more structure here, but rather than
retrace old work we chose to focus on the effect of other sources of
uncertainty.

\subsection{Other foregrounds}\label{subsec:otherfg}

\subsubsection{The Sun}

We find that it is important to include the quiet Sun in our
modelling, since this significantly affects our constraints on the
21-cm history. Although the Sun is a bright radio source, it is
compact rather than diffuse, so even if it lies at the centre of the
antenna beam its power is diluted by a factor of the solid angle
subtended by the Sun divided by the effective solid angle of the
antenna beam.  Fig.~\ref{fig:compare} shows its effective brightness
temperature (the brightness temperature of an object with the same
flux density but filling the beam) for this case. If the Sun lies away
from the centre of the beam, its power is suppressed even more.

For some of the observing time  of a lunar-orbiting antenna, the Sun
will be entirely occluded by the Moon. For the rest of the time, the
Sun's position in the antenna beam will vary and it will contribute
different amounts of power at different times, even if its intrinsic
luminosity remains perfectly steady. For that reason, in our
modelling, we assume that while the shape of the Sun's spectrum
remains the same, the overall contribution of the Sun (i.e.\ the
normalization of its integrated spectrum) will be different in each of
the eight sky areas we observe. Otherwise, our model for the Sun's
spectrum is similar to our model for the diffuse foreground spectrum,
i.e.\
\begin{equation}
\begin{split}
\log T^i_\mathrm{Sun}&=\log T^{i,\mathrm{Sun}}_0 +
a^\mathrm{Sun}_1\log (\nu/\nu_0) \\ & \quad +
a^\mathrm{Sun}_2[\log(\nu/\nu_0)]^2 +
a^\mathrm{Sun}_3[\log(\nu/\nu_0)]^3\ ,
\end{split}
\label{eqn:sunmodel}
\end{equation}
where $i=1,\ldots,8$ again labels the different sky areas, but $a_1$,
$a_2$ and $a_3$ do not carry an index, so that there are a total of 11
parameters to be fit. The input values of these parameters are derived
from a fit to the solar spectrum shown by \citet{ZAR04}, which yields
$a_1 = 1.9889$, $a_2 = -0.3529$ and $a_3 = 0.0407$. To our knowledge,
there are no observations which probe the variability of the solar
spectrum at the level of sensitivity required for our experiment, so
it is possible that in reality it has small variations in time, and
perhaps microbursts, contrary to our assumption. This may be tested in
the next few years with ground-based observatories such as the
Murchison Widefield Array (MWA) and the Long Wavelength Array. A full
3000 h dataset from our reference mission would allow the effect of
the Sun on the final constraints to be tested by using only those
observing times for which the Sun was occluded by the Moon.

Because we know the position of the Sun in the beam at all times
during our observations, and because ground-based observations of the
Sun with smaller beams than our proposed antenna may provide good
independent constraints on the solar spectrum, it is possible that
quite good priors may be placed on the parameters in
Equation~\eqref{eqn:sunmodel}. In Section~\ref{sec:disc} we consider
cases where these parameters are treated as being completely free and
fit only by the satellite data, and cases where good priors are placed
on them beforehand. The different shapes of the spectra of the diffuse
foregrounds and the Sun may lead one to worry that combinations of the
parameters of these sources may be degenerate with the parameters of
the 21-cm signal. In this case, what will help to disentangle the
21-cm signal from the foregrounds is that the former is identical
between different sky areas (because the antenna beam averages over
such an enormous volume of the high-redshift Universe) while the
latter varies spatially. It may also be possible to arrange that one
or two sky areas may be observed only when the Sun is occluded by the
Moon. Except where stated otherwise, our results below assume that two
out of the eight sky areas have always been observed when this is the
case. Were this not true, one may still be able to check that results
obtained excluding times during which the Sun is in view are
consistent with results from the full data set.  This assumption does
not make a significant difference to our results, but it does allow us
to examine how the presence or absence of the Sun in a given sky area
changes the correlation properties of the parameters in
Section~\ref{subsec:corr}.

During a solar burst, the radio power of the Sun can increase by
several orders of magnitude, and we would not anticipate using data
gathered during a solar burst for 21-cm work. The strength of the
bursts means, however, that they can be identified quite
straightforwardly, and data gathered during a burst can be excluded
unless the Sun is occluded by the Moon at the time. The excluded
periods would be short compared to the lifetime of the mission
\citep*[between a few seconds and $\sim\! 1$~hr; ][]{WIL63}, and there
are approximately tens of such bursts per year, depending on the phase
in the solar cycle \citep{GOP08}. Therefore we do not expect them to
significantly affect the sensitivity of the experiment.

\subsubsection{The Moon}

The Moon itself is a thermal radio source, with different radio
wavelengths probing its temperature at different depths. In our band,
the temperature is $\approx 220\ \mathrm{K}$
\citep[e.g.][]{SAL71,keihm1975}. Since the antenna will always be
pointing towards the sky rather than towards the lunar surface, the
Moon's contribution will be suppressed, though some radiation will
enter through the antenna's back- and sidelobes. Thus, its mean
effective contribution, shown in Fig.~\ref{fig:compare}, ends up being
around 20~K. We model the Moon's radiation using a single parameter --
its temperature -- and neglect any frequency dependence, or dependence
on phase in the lunar cycle, which is expected to be weak given the
depth probed by these long wavelengths. Clearly this may be an
oversimplification, but the data we have found do not yet seem precise
enough to suggest any specific, more sophisticated model, so this is
an area that may require further study.  If the emission from the Moon
is more complicated, it is possible that the modulation of its signal
as it enters more or less sensitive areas of the beam may help
disentangle it from the other sources, or that pointing the antenna
towards the Moon for some time may help constrain its emission at the
expense of a small amount of data collection time. An analysis would
then require the use of the full time series of spectra from the
satellite, rather than just the eight integrated spectra we look at
here, and is therefore beyond the scope of the current paper.

The Moon also reflects some of the radiation from the Galaxy and other
sources, and has a reflectivity of around 5--10 per cent
\citep[e.g.][]{DAV64}. We assume that this is constant with frequency,
so that the fraction of the incoming radiation which is reflected is a
single parameter in our model. The input value for this parameter is
chosen to be 10 per cent, though the true value is uncertain. This
reflected radiation is further suppressed, by a factor of around 10,
since it enters through the backlobe of the antenna, so that the
effective temperature of the reflected foregrounds, shown in
Fig.~\ref{fig:compare}, is around two orders of magnitude below that
entering the antenna directly from the front.

\subsubsection{Neglected contributions}

There are some other processes which one would expect to contribute to
the spectra but which are not explicitly included in the modelling. We
describe some of them here, and justify neglecting them in this
analysis.

Firstly, other planets, especially Jupiter, are known to be radio
sources at these frequencies. We expect Jupiter to have a
qualitatively similar effect on our results as the Sun, but taking
into account the small solid angle Jupiter subtends compared to the
size of our antenna beam it is fainter than the Sun by a factor of
around $10^{-4}$ \citep[using data from ][]{ZAR04}, and so we neglect
it here. Jupiter does experience intense bursts, but their spectrum
cuts off very sharply above around 40~MHz, so they are not expected to
intrude into our band.

Radio recombination lines \citep[RRLs; ][]{PET11} may comprise a
foreground which is not spectrally smooth. They are caused by
transitions of electrons between atomic energy levels with very large
principal quantum numbers, and can be seen either in absorption or
emission depending on frequency. So far, the only RRLs detected have
been from carbon atoms in our own Galaxy. The lines are narrow (around
10~kHz) and occur at known frequencies. The high resolution of our
unbinned spectra would therefore allow them to be detected (if
present) and removed while only discarding a very small fraction of
the data and having a negligible effect on our sensitivity. Indeed,
this is the main reason for requiring high spectral resolution. Since
we deal only with binned spectra in this paper, the RRLs are assumed
to have been excised before the rest of the analysis takes place, and
we do not include the effect of these excisions on the noise levels of
the binned spectra. It is possible that the integrated contribution of
RRLs from external galaxies (redshifted by various amounts) would
leave smoother low-level features in the spectrum. This contribution
has been estimated by \citet{petrovic2011} to be very small, however.
Moreover, it is unclear whether it could mimic the spectral features
expected in the 21-cm signal and, unlike the signal, it would not be
constant over the sky.

The relatively low altitude of the assumed lunar orbit means that the
spacecraft will encounter dust particles from the highly tenuous upper
layers of the lunar exosphere. Dust impacts at orbital speeds produce
puffs of plasma which generate an electrical response in the antenna
\citep{MEY85}, and therefore could be a source of noise. A calculation
of the noise power using a model for the height distribution of lunar
dust \citep{STU10} and for the surface area of the spacecraft,
however, gives the noise spectrum shown in Fig.~\ref{fig:compare}, at
least an order of magnitude below the thermal noise in a very deep
integration, and so we neglect dust impacts in this work.

Finally, we ignore noise or RFI reflections from other spacecraft
which may be visible from a low altitude orbit over the lunar far
side, such as others which orbit the Moon, or those positioned at the
Earth-Moon $\mathrm{L}_2$ point. Reflections of RFI from the sunshade
of the {\it James Webb Space Telescope}, for example, would be around
$10^{-7}$ of the thermal noise.

\subsection{The instrument}\label{subsec:inst}

Our model for instrumental effects on the measured spectrum caused by
the radiometer system (antenna, amplifiers, receiver and digital
spectrometer) is based on that used for EDGES \citep[see ][in
particular the supplementary information]{BOW10}. Internal calibration
is performed by switching the input between the antenna and
calibration loads. While a very precise absolute calibration is not
necessary, it is important that the relative calibration of different
channels is very accurate, since spectral smoothness at a level of one
part in $10^6$ is used during foreground removal, and a relative
calibration error could be confused with variations in the sky
spectrum. For wide-band systems such as {\it DARE} or EDGES, the
canonical internal calibration equation for a radiometer
\citep[e.g.][their eqn. 3]{bowman2008} is insufficient because the
impedance of the antenna varies strongly with frequency and is not
well-matched to the impedance of the receiver front-end amplifier
across the entire band as it would be for a narrow-band system.  The
primary result is that noise power emitted from the amplifier toward
the antenna (which is usually neglected in narrow-band systems) can be
reflected back into the receiver.  This noise can be correlated with
the downstream receiver noise producing constructive and destructive
interference as a function of frequency in the measured spectrum with
a period related to the electrical path length between the receiver
and the antenna. In order to account for this uncalibrated spectral
component, we use the noise wave propagation model
\citep{penfield1962,meys1978,weinreb1982,BOW10} to represent the
interaction of a noisy receiver amplifier and the antenna impedance.
Following \citet[his eqns. 5 and 6]{meys1978}, we have:
\begin{equation}
\begin{split}
T_\mathrm{ant}(\nu) &=T_\mathrm{a} + |\Gamma(\nu)|^2 T_\mathrm{b} +
2 T_\mathrm{c} |\Gamma(\nu)| \cos \left [ \beta(\nu) + \phi_\mathrm{c}  \right ] \\
& \quad + T_\mathrm{sky}(\nu) \left[1-|\Gamma(\nu)|^2\right]\ ,
\end{split}
\label{eqn:calib}
\end{equation}
where $T_\mathrm{a}$ is the standard noise from the output port of the
amplifier (usually called the receiver noise), $T_\mathrm{b}$ is the
noise directed toward the antenna from the input port of the
amplifier, and $T_\mathrm{c}$ is the amplitude of the correlated
components of $T_\mathrm{a}$ and $T_\mathrm{b}$ such that
$T_\mathrm{c} = \epsilon \sqrt{T_\mathrm{a} T_\mathrm{b}}$, where
$\epsilon$ is the amplitude of the correlation coefficient and the
parameter $\phi_\mathrm{c}$ in Eqn.~\eqref{eqn:calib} gives the phase
of the correlation.   $T_\mathrm{sky}$ is the total sky brightness
($T_\mathrm{21\mbox{-}cm} + T_\mathrm{FG} + T_\mathrm{Sun}+\ldots$)
convolved with the antenna beam and
$\Gamma(\nu)=|\Gamma(\nu)|\mathrm{e}^{i\beta(\nu)}$ is the reflection
coefficient of the antenna due to this impedance mismatch with the
receiver.

\begin{figure}
  \begin{center}
    \leavevmode
    \psfig{file=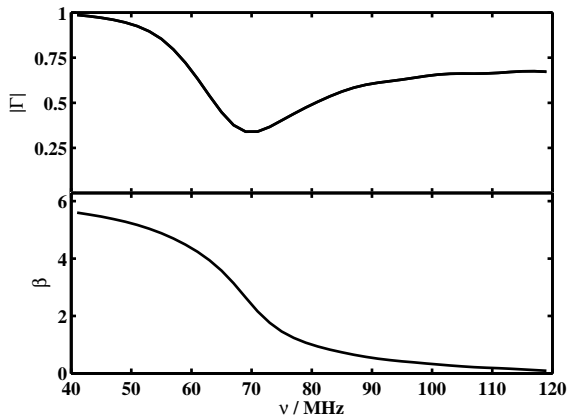,width=8cm}
    \caption{The reflection coefficient,
    $\Gamma(\nu)=|\Gamma|\mathrm{e}^{i\beta}$, due to the impedance
    mismatch between antenna and receiver for the simulated {\it DARE}
    radiometer system. The top panel shows $|\Gamma|$ and the bottom
    panel shows $\beta$ in radians. The sensitivity to the sky
    temperature scales as $1-|\Gamma |^2$.}\label{fig:gammabeta}
  \end{center}
\end{figure}

$T_\mathrm{a}$, $T_\mathrm{b}$, $T_\mathrm{c}$, $\phi_\mathrm{c}$,
$\Gamma$ and $\epsilon$ can be computed for a theoretical model of the
radiometer system, or estimated using measurements taken while the
satellite is on the ground. It is possible, however, that they cannot
be estimated to the required level of accuracy either way. Rather,
they may have to be estimated using the science data themselves. In
this paper, we assume that each of these parameters is constant as a
function of frequency, with the exception of the complex reflection
coefficient, $\Gamma$.   We further restrict the receiver noise
temperatures such that $T_\mathrm{a}\equiv T_\mathrm{b}$ always, and
we neglect $\phi_\mathrm{c}$ since it can be absorbed into the $\beta$
term with no loss of generality. We set the input value of the
receiver temperature to be 100~K and $\epsilon$ to be 0.1, retaining
each as a single free parameter in our model. As was done for the
antenna power pattern, $\Gamma(\nu)$ has been computed in an
electromagnetic model of the {\it DARE} radiometer system
\citep{burns2011}. The computed $\Gamma(\nu)$ is shown in
Fig.~\ref{fig:gammabeta}.
From Eqn.~\eqref{eqn:calib}, one can see that the sensitivity to the
sky temperature scales as $1-|\Gamma|^2$, and so
Fig.~\ref{fig:gammabeta} demonstrates that the system is most
sensitive at around 70~MHz, with the sensitivity tailing off somewhat
at high frequency and significantly at low frequency. By design the
frequency response is very smooth, to assist in foreground
subtraction.

We also used this smoothness to guide us in finding a suitable
parametrization for $\Gamma(\nu)$: we take as our parameters the ten
lowest-frequency coefficients of the discrete cosine transform (DCT)
of each of $|\Gamma(\nu)|$ and $\beta(\nu)$, giving us twenty
parameters in total. The DCT was chosen as a simple and efficient way
of modelling $\Gamma(\nu)$ as a sum of orthogonal functions. It was
chosen after some experimentation with various transforms as the one
which fit the simulated $\Gamma(\nu)$ to reasonable accuracy using a
small number of coefficients. Since the DCT is also simply a special
case of a real, discrete Fourier transform, it is very quick to
compute. $\Gamma(\nu)$ depends on the properties of the various
components of the antenna/receiver system, though, and so in future we
may hope to find a more physically motivated parametrization that
takes that into account. This would be useful if it could reduce the
number of parameters in the model or the degeneracies between them. To
find $\Gamma(\nu)$ given some set of parameters, we set all
higher-frequency coefficients to zero and then compute the two inverse
DCTs. Ten non-zero coefficients for each of $|\Gamma|$ and $\beta$ are
used since this is the least number that allows us to fit the small
ripples in the amplitude of the simulated reflection coefficient of
our reference experiment at high frequency. The coefficients obtained
thereby are the ones we use as the input to our modelling. When we
refer to simulating a hypothetical `perfect' instrument below, we take
this to mean that $\Gamma(\nu)$ is known to be identically zero.

\subsection{Thermal noise}

The noise on the spectrum, for a spectral bin of width $B$
observed for a time $t$, is given by the radiometer equation:
\begin{equation}
\sigma(\nu) = \frac{T_\mathrm{ant}(\nu)}{\sqrt{2Bt}}\ ,
\label{eqn:noise}
\end{equation}
where the factor of $\sqrt{2}$ in the denominator arises from the two
independent polarizations measured by a crossed dipole antenna. For an
observation near the centre of our band, where the coldest areas of
sky have $T_\mathrm{sky}\approx 1000\ \mathrm{K}$, and for $B=2\
\mathrm{MHz}$ and $t=375\ \mathrm{h}$ (corresponding to one of eight
sky areas, observed for one eighth of the total integration time of
3000 h) this gives a noise of $0.4\ \mathrm{mK}$ in each spectral
channel. This is roughly the level of noise above which \citet{PRI10}
found that constraints on the positions of turning points started to
be seriously degraded in their Fisher matrix analysis. For such long
integrations to be worthwhile, systematic sources of noise must also
be controlled to at least this level; we assume that this is the case
and do not attempt to model any additional systematic noise.
Systematics one might worry about include, for example, temperature
changes due to the spacecraft passing in and out of the shadow of the
Moon as it orbits, which could affect the noise properties of the
system, or leakage of noise from other components of the spacecraft
itself. The design of the reference experiment is intended to minimize
these effects: further details may be found in \citet{burns2011}.

\section{Model fitting}\label{sec:fitting}

For the fiducial case we consider, of a lunar-orbiting antenna
measuring the spectrum of eight independent sky regions, our model for
these eight spectra has 73 parameters:
\begin{itemize}
\item 6 for the 21-cm signal (frequency and temperature of three
  turning points);
\item 32 for the diffuse foregrounds (coefficients of a third-order
  polynomial in each of eight sky regions);
\item 11 for the Sun (a normalization parameter in each of eight
  different regions, plus three parameters describing the spectral
  shape);
\item 2 describing the Moon (one for its temperature and one for its
  reflectivity);
\item 22 describing the instrument (ten for the amplitude of the
    reflection coefficient, $|\Gamma(\nu)|$, ten for its phase,
    $\beta(\nu)$, and one each for $T_\mathrm{rcv}$ and
    $\epsilon$).
\end{itemize}
These parameters specify $T^i_\mathrm{ant}(\nu)$, where $i=1\ldots 8$
again runs over the different sky regions, and we can generate a
simulated realization of the experimental data by adding noise to
these spectra according to Eqn.~\eqref{eqn:noise}. Having different
sky regions in which the foregrounds are different but the 21-cm
signal is the same helps to break the degeneracy between the
foreground, signal and instrumental parameters.  We use eight regions
since averaging spectra into fewer sky regions would destroy
information, while splitting the sky into more areas would mean the
different spectra would not be independent, complicating the analysis.

Given the simulated noisy spectra, we find the best-fitting parameter
values, and confidence regions on these values, using a Markov Chain
Monte Carlo algorithm \citep[MCMC; e.g.\ ][and references
therein]{LEW02} we have implemented in \textsc{matlab}. This provides
an efficient way to explore a high-dimensional parameter space. Since
there are many good references on MCMC, we only provide enough of a
description of the technique here to establish some notation and allow
us to be more precise about our particular implementation.

We seek to map the posterior probability distribution
$P(\boldsymbol{\theta}|\boldsymbol{X})$ of the parameters of our
model. $P$ is considered to be a function of the vector of parameters
$\boldsymbol{\theta}$, with the vector $\boldsymbol{X}$, which
contains our simulated data, held fixed. The posterior is related to
the likelihood, $L(\boldsymbol{X}|\boldsymbol{\theta})$, by Bayes'
theorem,
\begin{equation}
P(\boldsymbol{\theta}|\boldsymbol{X})\propto
L(\boldsymbol{X}|\boldsymbol{\theta})P(\boldsymbol{\theta}),
\end{equation}
where $P(\boldsymbol{\theta})$ is the prior placed on the model
parameters.  For constant priors the likelihood gives us the posterior
probability, up to an arbitrary multiplicative constant.  Our goal is
to see how well the parameters $\boldsymbol{\theta}$ may be inferred
given an observed data set $\boldsymbol{X}$.

We assume that the noise in each frequency channel is Gaussian. Then,
if $T^i_\mathrm{ant}(\nu_j|\boldsymbol{\theta})$ is the predicted
antenna temperature in the $i^\mathrm{th}$ sky area in the
$j^\mathrm{th}$ frequency channel given a parameter set
$\boldsymbol{\theta}$, and $T^i_\mathrm{meas}(\nu_j)$ is the
`measured' temperature in this sky area and channel in a simulated
dataset, then the probability density of measuring this value is given
by
\begin{equation}
p_{ij} = \frac{1}{\sqrt{2\pi\sigma_i^2(\nu_j|\boldsymbol{\theta})}}
\mathrm{e}^{-[T^i_\mathrm{meas}(\nu_j)-T^i_\mathrm{ant}(\nu_j|\boldsymbol{\theta})]^2/2\sigma_i^2(\nu_j|\boldsymbol{\theta})}
\label{eqn:pij}
\end{equation}
where $\sigma_i(\nu_j|\boldsymbol{\theta})$ is the rms noise in the
channel, computed from $T^i_\mathrm{ant}(\nu_j|\boldsymbol{\theta})$,
the bandwidth and the integration time using
Eqn.~\eqref{eqn:noise}. Then, assuming that each sky area and
frequency channel is independent, the likelihood is given simply by
\begin{equation}
L(\boldsymbol{T}_\mathrm{meas}|\boldsymbol{\theta}) = \prod_{i=1}^{\
n_\mathrm{areas\phantom{q}}}\prod_{j=1}^{n_\mathrm{freq}}p_{ij} \ ,
\label{eqn:like}
\end{equation}
where $\boldsymbol{T}_\mathrm{meas}$ is a vector containing
$T^i_\mathrm{meas}(\nu_j)$ for all $i$ and $j$. Usually it is
computationally simpler to work with $\log(L)$, for which the double
product becomes a double sum.

We use the Metropolis-Hastings algorithm \citep{hastings1970} to
generate a sequence of random samples from the posterior distribution;
this sequence of samples is a chain. To see how the algorithm works,
suppose the chain is at the position $\boldsymbol{\theta}_n$ in
parameter space. We randomly generate another parameter vector,
$\boldsymbol{\theta}_{n+1}$, according to the `proposal density'
$q(\boldsymbol{\theta}_n,\boldsymbol{\theta}_{n+1})$. This vector is
accepted as the next link in the chain with a probability
\begin{equation}
\alpha(\boldsymbol{\theta}_n,\boldsymbol{\theta}_{n+1})=\min\left\{
1,\frac{P(\boldsymbol{\theta}_{n+1}|\boldsymbol{X})}{P(\boldsymbol{\theta}_n|\boldsymbol{X})}\right\}
\ ,
\label{eqn:accept}
\end{equation}
if $q$ is symmetric, which is true for the proposal densities we use.
Over time the chain will explore the full parameter space with
statistical properties that allow a set of unbiased and random samples
to be extracted.

Clearly the position of successive samples is correlated.  To reduce
this correlation and obtain approximately independent samples, we
`thin' the chain, retaining only one out of every $n_\mathrm{thin}$
samples. The results we show here use $n_\mathrm{thin}=50$, which
allows us to run the chains long enough to reach convergence without
having to store an extremely large number of samples. The first
$n_\mathrm{burn\mbox{-}in}$ thinned samples are discarded to ensure we
only use samples from the equilibrium distribution; we find
$n_\mathrm{burn\mbox{-}in}=10^4$ to be sufficient.

The choice of the proposal density, $q$, strongly affects the
computational efficiency of the algorithm. A $q$ which is too broad
makes the acceptance ratio, $\langle\alpha\rangle$, very small,
meaning we have to draw from $q$ and compute $L$ many times to obtain
each new link in the chain. A $q$ which is too narrow forces us to
take only tiny steps in parameter space, preventing us from mapping
its interesting regions in a reasonable amount of time. To avoid
either of these scenarios, we automate the choice of $q$. This
requires us first to estimate the parameter covariance matrix,
$\mathbfss{C}$. For the first $2n_\mathrm{burn\mbox{-}in}$ thinned
samples following burn-in, we compute an estimate of $\mathbfss{C}$
from an estimate of the Hessian of the posterior, using the
\textsc{derivest} package. For subsequent samples, we instead estimate
$\mathbfss{C}$ directly from the cloud of existing samples, which we
have found to be more robust. The covariance matrix, being costly to
compute, is recalculated only after every $10^5$ evaluations of the
posterior. Because the proposal distribution changes during a run,
this means that the chains are no longer strictly Markov. During the
preparation of this manuscript, however, it came to our attention that
this approach is very similar to the `Adaptive Metropolis' algorithm
of \citet*{haario2001}, who prove that the chain none the less has the
desired ergodicity properties and converges correctly to the
equilibrium solution.

Having found $\mathbfss{C}$, we proceed to find a basis of parameter
space in which it is diagonal, and denote by
$\tilde{\boldsymbol{\theta}}$ the vector of parameters in this new
basis. We choose a random subset of these parameters, $\{
\tilde{\theta}_{i_1},\tilde{\theta}_{i_2},\ldots,\tilde{\theta}_{i_{n_\mathrm{vary}}}
\}$ to vary at each step, where $n_\mathrm{vary}$ is a numerical
parameter we may choose. Taking $n_\mathrm{vary}=1$ typically gives us
acceptance ratios of around 70 per cent. The proposal distribution $q$
for this subset of the (transformed) parameters is then taken to be a
multivariate Gaussian distribution, with a diagonal covariance matrix
equal to that of the full parameter set, but keeping only the rows and
columns numbered $i_1,i_2,\ldots,i_{n_\mathrm{vary}}$. This choice
appears to perform well in practice.

We obtain around $10^5$ thinned samples in a few hours on a 2.3 GHz
AMD Opteron processor. By running eight different chains we can apply
the convergence test of \citet{GEL92}, which confirms the impression
from a single chain that using this many samples is sufficient for
good convergence.

\begin{figure}
  \begin{center}
    \leavevmode
    \psfig{file=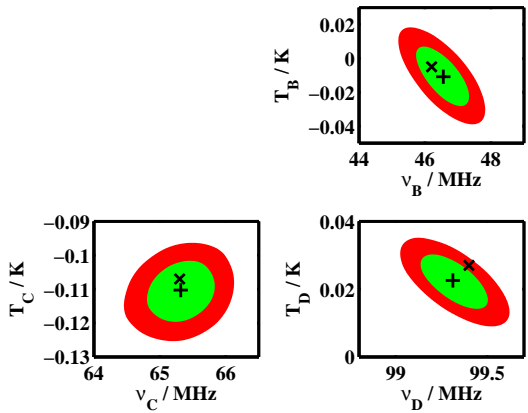,width=8cm}
    \caption{Confidence regions on turning points B, C and D of the
    cosmological signal, assuming a perfect instrument observing eight
    areas of the sky for a total of 3000 h. The 68 per
    cent confidence region is in green, and the 95 per cent confidence
    region is in red. For each turning point, the frequency of the
    turning point is on the $x$ axis and the brightness temperature on
    the $y$ axis. The `+' shows the best-fitting parameter values,
    while the `$\times$' shows the parameter values which were used as
    inputs to the simulation.}\label{fig:cont_perf}
  \end{center}
\end{figure}

An example of marginalized parameter distributions obtained using this
method for a particular noise realization is shown in
Fig.~\ref{fig:cont_perf}. In this case we assume a perfect instrument
($\Gamma(\nu)\equiv 0$) observing eight sky areas for a total of 3000
h, i.e.\ 375 h per sky area. Each panel shows the joint distribution
of the frequency and temperature of one of the turning points of the
21-cm signal. We also show the position of the input value of the
parameters and the parameter values with the highest posterior
probability. The difference between these gives a sense of how
robustly the parameters may be recovered or whether the recovered
parameters may be biased in some way by the foreground removal
process.  We will use this format for displaying most of our results.

For this particular noise realization, the input parameter values for
two turning points lie within the 68 per cent confidence region, while
those for the third lie just outside its border, suggesting that the
parameters have been recovered without significant bias, even for the
most difficult turning point at low frequency (turning point B).
Further, the size of the confidence regions is small and, thus, very
promising, allowing the three turning points to be distinguished
clearly from one another.

\section{Results and discussion}\label{sec:disc}

\begin{figure}
  \begin{center}
    \leavevmode
    \psfig{file=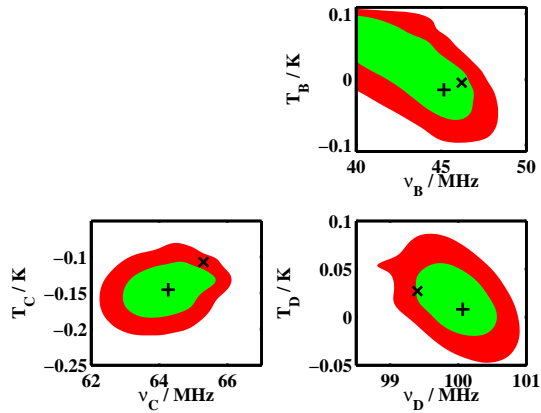,width=8cm}
    \caption{Confidence regions on turning points B, C and D of the
    cosmological signal, assuming a realistic instrument, and that we
    have no prior information on the parameters of the instrument
    model and the solar spectrum. Eight sky regions are observed for a
    total of 3000 h. Colours and symbols are as for
    Fig.~\ref{fig:cont_perf}, but for this and all subsequent figures
    the axis scales vary.}\label{fig:cont_withall_loose}
  \end{center}
\end{figure}

Having looked at signal recovery for a perfect instrument, we now move
on to a more realistic case based on the simulated properties of the
proposed {\it DARE} satellite. We start by assuming that there is no
meaningful prior information on any of the parameters, so that they
are constrained only by the satellite science data. The confidence
regions for this case, for a single random noise realization assuming
a total of 3000 h of integration time (375 h per sky area), are shown
in Fig.~\ref{fig:cont_withall_loose}. It is easy to see, noting the
difference in axis scale between Figs.~\ref{fig:cont_perf}
and~\ref{fig:cont_withall_loose}, that the parameter constraints are
significantly degraded. The frequency of turning point C, for example,
is found with an error of around $\pm 1\ \mathrm{MHz}$, rather than
$\pm 0.5\ \mathrm{MHz}$ from a perfect instrument. The best-fitting
values of all the parameters are somewhat offset from the true values,
but the error appears to be consistent with the confidence regions
estimated from MCMC. Turning point B is worst affected: the 68 per
cent confidence region spans a range of well over 100 mK in
temperature, and extends in frequency to below the bottom end of the
range (40 MHz), where we have truncated the scale of the plot. The
constraint on its frequency is therefore very model-dependent, and is
probably best viewed as an upper limit, ruling out a turning point
above $\sim 48\ \mathrm{MHz}$ with 95 per cent confidence (this upper
limit is properly computed from the fully marginalized,
one-dimensional probability distribution of $\nu_\mathrm{B}$; see
Table~\ref{tab:conf_int}).

The temperature of the other turning points is not as well determined
as their frequency, in the sense of how constraining the limits would
be for models of the dark ages, but some measurements are still
obtained. For all three turning points, the temperature is slightly
underestimated. This is simply because the temperature errors are
correlated across the band rather than because of some bias in the
method: we examine the correlations further in
Section~\ref{subsec:corr}. The correlation arises because of the
difficulty of measuring the overall normalization of the signal (as
opposed to its spectral variation) in the presence of the strong
foregrounds.

Even though the parameter constraints may look weak compared to the
case for the perfect instrument of Fig.~\ref{fig:cont_perf}, the
foreground parameters are very well constrained. For example, the
spectral index in one of the sky regions at $80\ \mathrm{MHz}$ is
found to be $-2.350898\pm 0.000042$ (2-$\sigma$ errors), compared to a
true value of $-2.350903$. This illustrates the dynamic range required
for such an experiment. The Sun, being a weaker source, does not have
its spectral index determined quite so well: we find a value of
$1.991\pm 0.011$ at $80\ \mathrm{MHz}$, with the true value being
$1.989$.

\begin{figure}
  \begin{center}
    \leavevmode
    \psfig{file=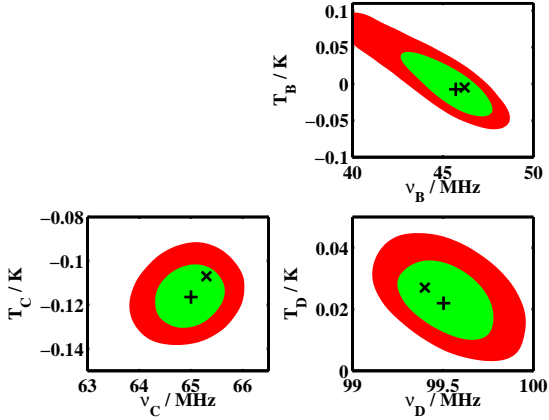,width=8cm}
    \caption{Confidence regions on turning points B, C and D of the
    cosmological signal, assuming a realistic instrument observing
    eight sky regions for a total of 3000 h, but with tight priors on
    the parameters of the instrument model, the solar spectrum and the
    properties of the Moon. We assume here, as throughout, that there
    is no prior information on the parameters of the diffuse
    foregrounds or the 21-cm signal itself. Colours and symbols are as
    for Fig.~\ref{fig:cont_perf}.}\label{fig:cont_withall_tight}
  \end{center}
\end{figure}

Fig.~\ref{fig:cont_withall_tight} shows how the constraints are
improved if we impose tight, Gaussian priors on the parameters of the
non-diffuse foregrounds and the instrument, again assuming 3000 h of
observation. $T_0^{i,\mathrm{Sun}}$ is assumed to be known to 0.1 per
cent for all $i$, as are the temperature and reflectivity of the Moon,
while $\{a_1^\mathrm{Sun},a_2^\mathrm{Sun},a_3^\mathrm{Sun}\}$ are
known with an error of $\pm 0.001$. The coefficients of $\Gamma(\nu)$
are known to one part in $10^6$ (i.e. almost perfectly), while
$T_\mathrm{rcv}$ and $\epsilon$ are known to 0.1 per cent. The
assumption that the reflection coefficient is known to one part in
$10^6$ is well beyond typical expectations at present, and is thus an
optimistic prior. Most antennas are characterized at the 1 per cent
level today, but devices designed to make accurate impedance
measurements are stated in their specifications to perform to an
accuracy of $<0.1$ per cent, and it is reasonable that this level
could be achieved. This topic is being actively worked on with EDGES,
the closest current analogue to {\it DARE}, in the field and the
laboratory, with a target of achieving an accuracy of one part in
$10^4$. Furthermore, it should be possible to treat the unknown
aspects of the reflection coefficient with more physically motivated
models than the DCT, which would help to reduce the effective degrees
of freedom and so approach the desired accuracy. Our priors on
$T_\mathrm{rcv}$ and $\epsilon$ are more plausible, and they could
well be measured in the lab to this level before launch.

Under these conditions, the parameter constraints approach more
closely those for the perfect instrument of Fig.~\ref{fig:cont_perf}.
The main exception is that it becomes harder to rule out turning point
B lying at a much lower frequency and higher temperature. A good
measurement can only be found at 68 per cent confidence. The 95 per
cent confidence region extends outside the band for which we have
data, and any inferences about the properties of the signal in that
region depend strongly on the assumed signal model. The shape of the
confidence region suggests that our data actually tell us the
amplitude and slope of the signal at low frequency, and that turning
point B lies somewhere on a curve consistent with that amplitude and
slope within the errors.

Since the instrumental frequency response and the non-diffuse
foregrounds are well known, the weak constraint on turning point B
compared to the perfect instrument must occur because of the reduced
sensitivity at low frequencies, caused by the large value of
$|\Gamma(\nu)|$ there. To find the position of turning point B
precisely, it may be necessary to have an instrument with better
sensitivity at low frequency, and possibly a lower minimum frequency.
This is difficult to achieve (see the steep drop in sensitivity at low
frequencies in Fig.~\ref{fig:gammabeta}), though one possible route
would be a larger antenna and ground screen, which may be awkward and
expensive for a satellite mission. Even then, it is hard to design an
antenna which can cover a frequency range which is more than a factor
of $\sim\! 3$ without (for example) the antenna changing mode at the
top end of the frequency range and introducing frequency structure
into the response. The large uncertainty in current theoretical models
of the signal means that an instrument with a range of, say,
$35$--$105\ \mathrm{MHz}$ would run the risk of missing out entirely
on turning point D, which we would otherwise hope to constrain quite
precisely. Figure~\ref{fig:cont_withall_tight} suggests we are close
enough to a measurement of turning point B that this may be possible
with some smaller tweak to the design without having to change the
current {\it DARE} frequency range.

\begin{figure}
  \begin{center}
    \leavevmode
    \psfig{file=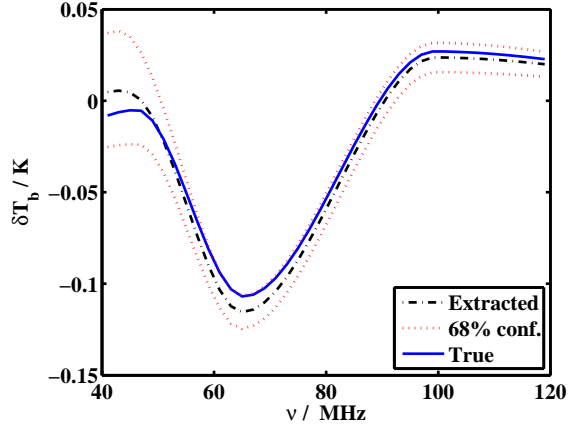,width=8cm}
    \caption{The mean extracted signal (dot-dashed black line)
    and 68 per cent confidence limits on this mean (dashed red
    lines) are compared to the `true' signal constructed from
    the input parameters to the simulation (solid blue line).
    This plot assumes an instrument observing eight areas of the
    sky for a total of 3000 h, with tight priors on the
    parameters concerning the instrument, the Sun and the Moon,
    as in
    Fig.~\ref{fig:cont_withall_tight}.}\label{fig:path_alltight}
  \end{center}
\end{figure}

We show how the parameter constraints of
Fig.~\ref{fig:cont_withall_tight} translate into constraints on the
shape of the 21-cm signal in Fig.~\ref{fig:path_alltight}. Here we
plot the true signal, the mean over all the samples of the extracted
signal at each frequency, and a 68 per cent confidence interval around
this mean. The shape of the signal is recovered quite well, but the
frequency of the turning points seems, visually, to be recovered more
accurately than the temperature. The absolute normalization of the
curve is difficult to determine.

The width of the error bars is larger at the lower end of the
frequency range than at the upper end, but not to the extent that
would be expected if one were simply to use the rms thermal noise at
each frequency to determine the error bar, since the sky temperature
in the lowest frequency channel is $>10$ times that in the highest
frequency channel, and this is the most important contributor to the
thermal noise. Instead, the errors across the whole band are highly
correlated, since the shape of the signal is reconstructed only from
the six parameters giving the position of the turning points.

\begin{figure*}
\vspace*{0.5cm}
\begin{center}
\psfig{file=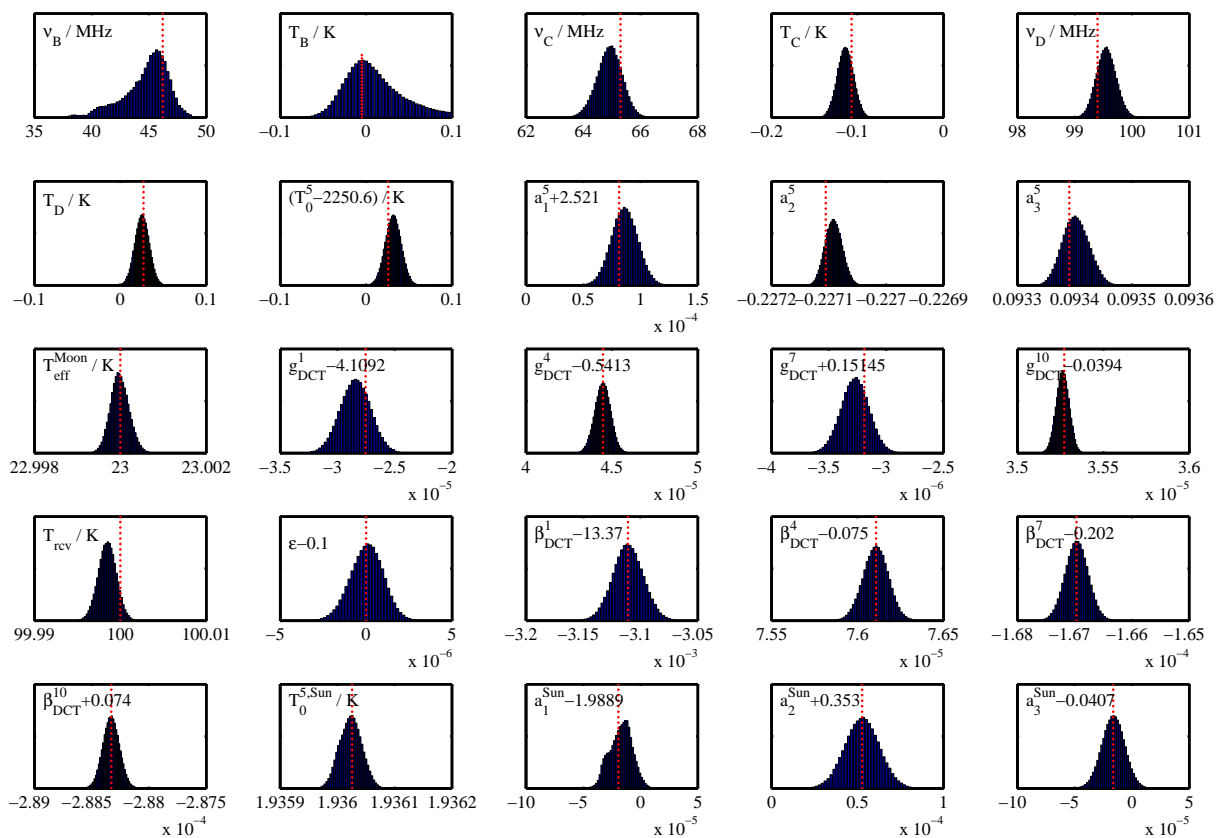,width=17cm} \caption{1D
marginalized distributions of the parameters for the case when we have
tight priors on the properties of the instrument, the Sun and the
Moon, as in Figs.~\ref{fig:cont_withall_tight} and
\ref{fig:path_alltight}. The vertical, red, dashed line shows the
input value of the parameter. The names of the parameters given in
each panel are as in the text, except that the discrete cosine
transform coefficients of $|\Gamma|$ and $\beta$ are labelled
$g_\mathrm{DCT}^i$ and $\beta_\mathrm{DCT}^i$
respectively.}\label{fig:1d_dists}
\end{center}
\end{figure*}

Before moving on from the case where we have good prior information on
the properties Sun, Moon and instrument, we illustrate the errors on
individual parameters which can be achieved in this case by showing
the marginalized distributions of a subset of them in
Fig.~\ref{fig:1d_dists}. It is also reassuring to be able to check
that the distributions seem fairly smooth and well behaved.
Complicated, multimodal distributions (or, for example, strongly
curving degeneracies between different parameters) would be awkward
for the sampler we have implemented here, and might require a more
sophisticated method to sample them efficiently.

\begin{figure}
  \begin{center}
    \leavevmode
    \psfig{file=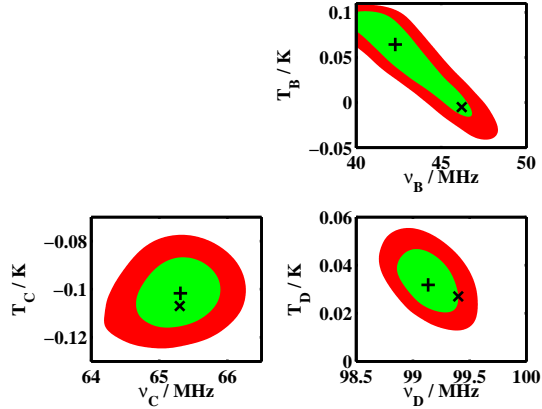,width=8cm}
    \caption{Confidence regions on turning points B, C and D of the
    cosmological signal, assuming a realistic instrument observing
    eight sky regions for a total of 3000 h, but with tight priors on
    the parameters of the instrument model. Colours and symbols are as
    for Fig.~\ref{fig:cont_perf}.}\label{fig:cont_withall_tightinst}
  \end{center}
\end{figure}

We now wish to consider whether the improvement between
Fig.~\ref{fig:cont_withall_loose} and
Fig.~\ref{fig:cont_withall_tight} comes from our better knowledge of
the non-diffuse foregrounds (in particular the Sun) or of the
instrument. To this end, in Fig.~\ref{fig:cont_withall_tightinst} we
show results obtained using the tight priors on $\Gamma(\nu)$ given
above, but reverting to weak priors on the spectral parameters of the
Sun.

The results for turning points C and D are almost as good as for the
previous case, showing that superb knowledge of the instrument is the
most important factor in extracting the 21-cm signal accurately,
though constraints on turning point B are noticeably degraded.
Foreground parameters are also measured more precisely than for the
case of Fig.~\ref{fig:cont_withall_loose}: for example, the error on
the spectral index of the Sun at 80 MHz is reduced by a factor of
about six. If the instrumental calibration can be improved by using
the spectra at full time and frequency resolution, or by introducing
extra mechanisms for internal calibration, then this would clearly be
very desirable, and should be the subject of further study.

\begin{figure}
  \begin{center}
    \leavevmode
    \psfig{file=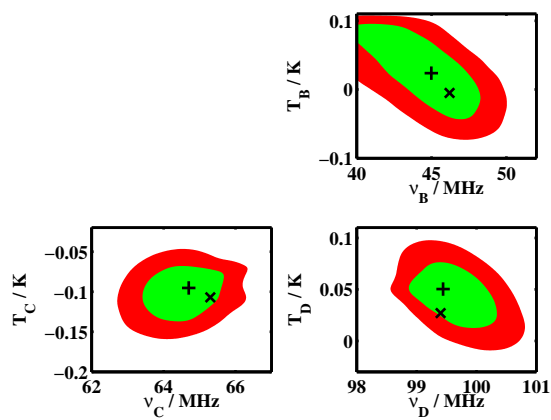,width=8cm}
    \caption{Confidence regions on turning points B, C and D of the
    cosmological signal, assuming a realistic instrument observing
    eight sky regions for a total of 3000 h, and with tight priors on
    the parameters of the solar spectrum and the Moon. Colours and
    symbols are as for
    Fig.~\ref{fig:cont_perf}.}\label{fig:cont_withall_tightnoninst}
  \end{center}
\end{figure}

By contrast with Fig.~\ref{fig:cont_withall_tightinst},
Fig.~\ref{fig:cont_withall_tightnoninst} shows the confidence regions
we derive when we assume tight priors on the spectrum of the Sun
(obtained perhaps by ground-based observations), but relax the priors
on the coefficients of the instrumental response to their original
size.  The constraints on the signal parameters are improved only a
little over those of Fig.~\ref{fig:cont_withall_loose}, with the
overall temperature normalization being especially hard to recover.
None the less, external constraints on the solar spectrum would be
valuable as a consistency check.

\begin{figure}
  \begin{center}
    \leavevmode
    \psfig{file=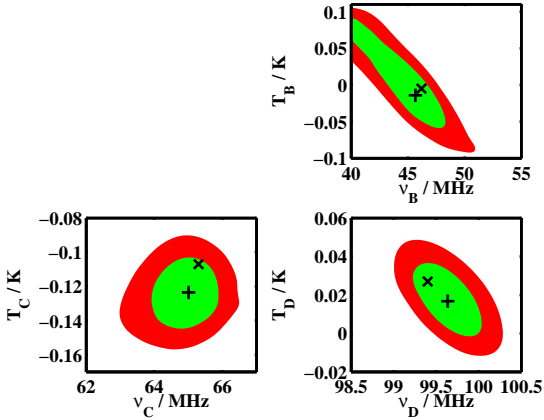,width=8cm}
    \caption{Confidence regions on turning points B, C and D of the
    cosmological signal, assuming a realistic instrument observing
    eight sky regions for a total of 1000 h, and with tight priors on
    the parameters of the instrument model, the solar spectrum and the
    Moon. Colours and symbols are as for
    Fig.~\ref{fig:cont_perf}. This figure should be compared to
    Fig.~\ref{fig:cont_withall_tight}, which makes the same
    assumptions and differs only in the amount of integration
    time.}\label{fig:cont_withall_1000}
  \end{center}
\end{figure}

Finally, we look at the effect of changing the available integration
time. Results so far have used 3000 h of data; for
Fig.~\ref{fig:cont_withall_1000} we assume instead only 1000 h of
data, as may occur if the satellite is able to observe for only one
year. Otherwise, the assumptions are the same as for
Fig.~\ref{fig:cont_withall_tight}, i.e.\ tight priors on both the
instrument and the non-diffuse foregrounds are assumed.

\begin{table*}
\begin{minipage}{175mm}
 \caption{95 per cent confidence intervals (or, in some cases, upper
 and lower limits) on the frequency, redshift and temperature of
 turning points B, C and D for the various sets of assumptions we have
 considered. The first column gives a brief description of each
 simulation, while the second shows which figures were plotted using
 data from that simulation. The remaining columns show the
 constraints. The first row shows the true input values of the
 parameters, for comparison. All simulations assume 3000 h of
 observation, unless otherwise stated. A fuller description of each
 model is given in the relevant figure captions.}
 \label{tab:conf_int}
 \begin{tabular}{@{}lcccccccccc}
  \hline
   & & \multicolumn{3}{c}{Turning
  point B} & \multicolumn{3}{c}{Turning point C} &
  \multicolumn{3}{c}{Turning point D} \\
  Description & Figures & $\nu / \mathrm{MHz} $ & $z$ &
   $T / \mathrm{mK}$ & $\nu / \mathrm{MHz}$ & $z$ & $T / \mathrm{mK}$
   & $\nu / \mathrm{MHz}$ & $z$ & $T / \mathrm{mK}$ \\
  \hline
  True input values & - & 46.2 & 29.7 & $-5$ & 65.3 & 20.8 & $-107$ &
   99.4 & 13.29 & 27 \\
  Perfect instrument & \ref{fig:cont_perf} & $46.6^{+0.9}_{-1.1}$ &
   $29.5^{+0.7}_{-0.6}$ & $-10^{+22}_{-22}$ & $65.3^{+0.7}_{-0.6}$ &
   $20.7^{+0.3}_{-0.2}$ & $-111^{+11}_{-11}$ & $99.3^{+0.3}_{-0.2}$ &
   $13.30^{+0.03}_{-0.03}$ & $23^{+9}_{-12}$ \\
  No prior information &
   \ref{fig:cont_withall_loose},\ref{fig:scaled_covmat} & $<47.6$ & $>28.8$ & $55^{+45}_{-112}$ & $64.3^{+1.3}_{-1.6}$ & $21.1^{+0.6}_{-0.5}$ & $-141^{+43}_{-55}$ & $100.0^{+0.7}_{-0.8}$ & $13.21^{+0.11}_{-0.10}$ & $8^{+61}_{-41}$ \\
  All tight priors &
   \ref{fig:cont_withall_tight},\ref{fig:path_alltight},\ref{fig:1d_dists} & $45.6^{+2.4}_{-5.2}$ & $30.1^{+4.1}_{-1.5}$ & $-7^{+84}_{-42}$ & $65.0^{+0.8}_{-0.9}$ & $20.9^{+0.3}_{-0.3}$ & $-116^{+18}_{-17}$ & $99.5^{+0.4}_{-0.3}$ & $13.27^{+0.05}_{-0.05}$ & $23^{+17}_{-15}$ \\
  Tight inst.\ priors & \ref{fig:cont_withall_tightinst} & $42.6^{+4.8}_{-3.1}$ & $32.3^{+2.7}_{-2.3}$ & $95^{+6}_{-111}$ & $65.3^{+0.7}_{-0.8}$ & $20.7^{+0.3}_{-0.2}$ & $-102^{+18}_{-17}$ & $99.1^{+0.4}_{-0.3}$ & $13.33^{+0.05}_{-0.05}$ & $32^{+18}_{-14}$ \\
  Tight non-inst.\ priors & \ref{fig:cont_withall_tightnoninst} & $<49.0$ & $>28.0$ & $25^{+75}_{-65}$ & $64.7^{+1.3}_{-1.6}$ & $21.0^{+0.5}_{-0.4}$ & $-94^{+34}_{-52}$ & $99.6^{+1.0}_{-0.8}$ & $13.26^{+0.11}_{-0.13}$ & $51^{+35}_{-48}$ \\
  1000 h integration & \ref{fig:cont_withall_1000} & $<48.9$ & $>28.1$
   & $-11^{+108}_{-51}$ & $65.0^{+1.1}_{-1.5}$ & $20.9^{+0.4}_{-0.4}$
   & $-125^{+26}_{-24}$ & $99.6^{+0.5}_{-0.4}$ &
   $13.26^{+0.07}_{-0.08}$ & $17^{+24}_{-22}$ \\
  10000 h integration & \ref{fig:cont_withall_10000} &
   $46.9^{+1.5}_{-4.4}$ & $29.3^{+3.2}_{-1.0}$ & $-9^{+71}_{-22}$ &
   $65.4^{+0.7}_{-0.5}$ & $20.7^{+0.2}_{-0.2}$ & $-98^{+16}_{-11}$ &
   $99.4^{+0.2}_{-0.3}$ & $13.30^{+0.03}_{-0.03}$ & $35^{+11}_{-11}$ \\
  \hline
 \end{tabular}
\end{minipage}
\end{table*}

The effect is as one might expect, with confidence regions on the
parameters being enlarged somewhat. Turning points C and D can still
be localized: a single year of data from our reference experiment
could yield a detection of the first astrophysical sources of heating
in the Universe, and the start of the epoch of reionization. It
becomes impossible to obtain anything other than an upper limit on the
frequency of turning point B, however: the sensitivity at the low
frequencies is simply not sufficient for a clear measurement of its
position.

\begin{figure}
  \begin{center}
    \leavevmode
    \psfig{file=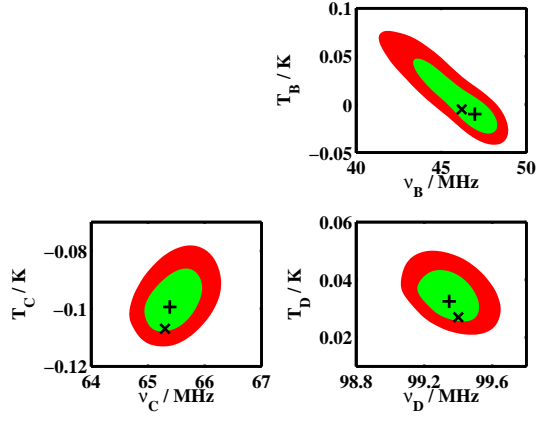,width=8cm}
    \caption{Confidence regions on turning points B, C and D of the
    cosmological signal, assuming a realistic instrument observing
    eight sky regions for a total of 10000 h, and with tight priors on
    the parameters of the instrument model, the solar spectrum and the
    Moon. Colours and symbols are as for
    Fig.~\ref{fig:cont_perf}. This figure should be compared to
    Figs.~\ref{fig:cont_withall_tight} and
    \ref{fig:cont_withall_1000}, which make the same assumptions and
    differ only in the amount of integration
    time.}\label{fig:cont_withall_10000}
  \end{center}
\end{figure}

To make sure that a realistic instrument can find a firm, 2-$\sigma$
detection of the frequency of turning point B given sufficient
integration time, we show results for 10000 h of observation in
Fig.~\ref{fig:cont_withall_10000}. In this case the 2-$\sigma$
contours do indeed close above $40\ \mathrm{MHz}$, though there is
still a significant degeneracy between the frequency and temperature
of turning point B. The positions of turning points C and D are
measured with improved accuracy compared to our baseline case, though
further study of such deep integrations may need the possible
systematics to be considered more carefully. An integration of this
length would be challenging from space, needing either a mission of
long duration or a very high observing efficiency (possibly both). It
is likely that the requisite noise level in the vicinity of turning
point B can be achieved more easily by modifications to the design of
the spacecraft or radiometer system. Better constraints on turning
point B might also come by extending the frequency coverage to lower
frequencies.

We summarize the constraints on the parameters of the signal for all
the different assumptions we have considered in
Table~\ref{tab:conf_int}. Here, we show 95 per cent confidence
intervals (or, in some cases, upper or lower limits) on the frequency,
redshift and temperature of the turning points, and record the figures
for which each set of assumptions was used.

\subsection{Correlation between parameters}\label{subsec:corr}

\begin{figure*}
\vspace*{0.5cm}
  \begin{center}
    \psfig{file=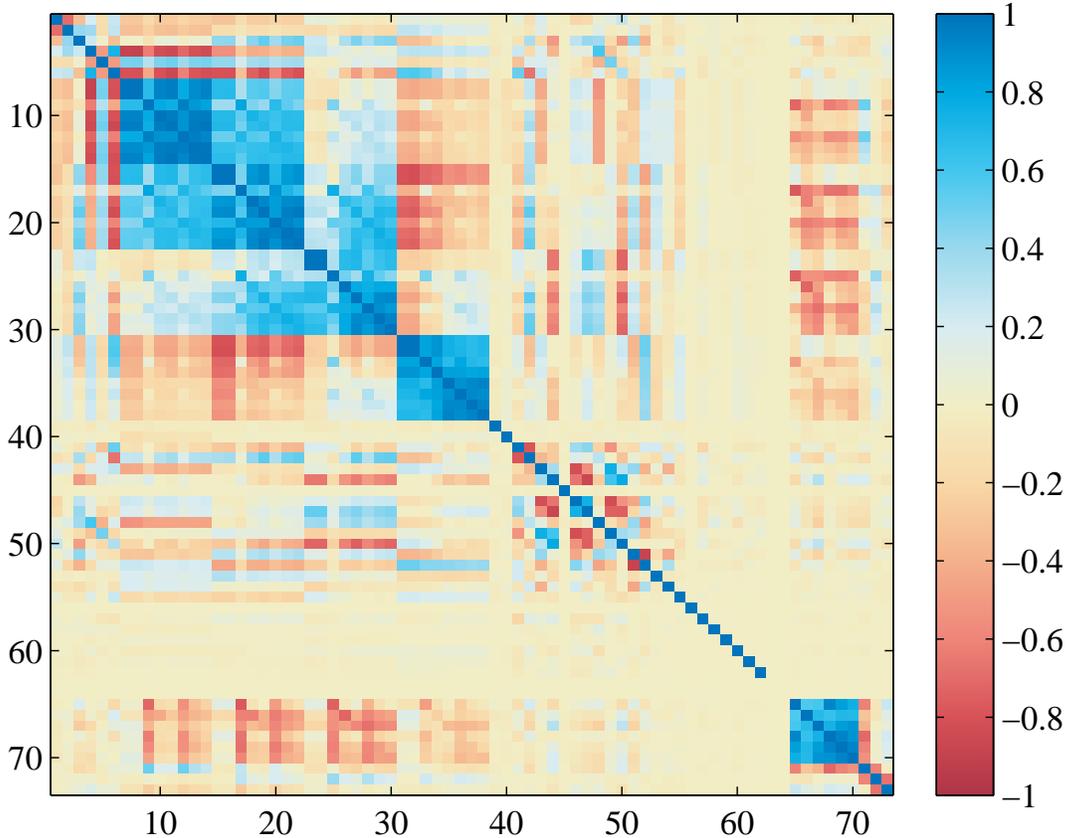,width=17cm}
    \caption{The scaled covariance matrix of all the parameters of the
    model, for a realistic instrument and assuming no meaningful prior
    information on the parameter values, as for
    Fig.~\ref{fig:cont_withall_loose}. By the `scaled' covariance
    matrix, we mean the that each pixel shows the correlation
    coefficient between two parameters, where a value of zero implies
    no correlation, and a value of 1 ($-1$) means perfect
    (anti-)correlation. The 1s on the diagonal come about because
    each variable is perfectly correlated with itself. The order of
    the parameters is given in Table~\ref{tab:parnumbering}. Note that
    parameters 63 ($T_0^{1,\mathrm{Sun}}$) and 64
    ($T_0^{2,\mathrm{Sun}}$) are set to be identically zero since the
    Sun is assumed to be occluded by the Moon in sky regions 1 and 2,
    which accounts for the obvious stripe at this
    position. This figure is best viewed in colour, to make the
    difference between correlations (blue) and anticorrelations (red)
    more clear.}\label{fig:scaled_covmat}
  \end{center}
\end{figure*}
\begin{table}
 \caption{Numbering of the rows of the scaled covariance matrix in Fig.~\ref{fig:scaled_covmat}}
 \label{tab:parnumbering}
 \begin{tabular}{@{}ll}
  \hline
  Row/column number & Parameter description \\
  \hline
  1--2 & Frequency and temperature of turning point B \\
  3--4 & Frequency and temperature of turning point C \\
  5--6 & Frequency and temperature of turning point D \\
  7--14 & $T_0^i$, i=1,\ldots{},8 \\
  15--22 & $a_1^i$, i=1,\ldots{},8 \\
  23--30 & $a_2^i$, i=1,\ldots{},8 \\
  31--38 & $a_3^i$, i=1,\ldots{},8 \\
  39 & Effective temperature of the Moon \\
  40 & Reflectivity of the Moon \\
  41--50 & Discrete cosine transform coefficients of $|\Gamma(\nu)|$ \\
  51--52 & $T_\mathrm{rcv}$ and $\epsilon$ \\
  53--62 & Discrete cosine transform coefficients of $\beta(\nu)$ \\
  63--70 & $T_0^{i,\mathrm{Sun}}$, i=1,\ldots{},8 \\
  71--73 & $a_1^\mathrm{Sun}$, $a_2^\mathrm{Sun}$ and
  $a_3^\mathrm{Sun}$ \\
  \hline
 \end{tabular}
\end{table}

The contour plots we have shown allow one to see clearly if the
inferred frequency and temperature of a given turning point are
correlated, or in other words if there is a degeneracy between these
two parameters. The frequency, $\nu_\mathrm{B}$, and temperature,
$T_\mathrm{B}$, of turning point B, for example, are clearly
anticorrelated in all our figures. Such correlations may exist between
all our parameters, and allow one to pick out possible degeneracies.
Therefore, in Fig.~\ref{fig:scaled_covmat}, we show a scaled version
of the covariance matrix of the parameters, such that a value of 1
($-1$) in pixel $\{i,j\}$ indicates that the value of parameters $i$
and $j$ in the MCMC samples is perfectly \mbox{(anti-)}correlated,
with a value of zero indicating no correlation. The key to the
numbering of the rows and columns of the image is given in
Table~\ref{tab:parnumbering}.

As expected, parameters 1 and 2 ($\nu_\mathrm{B}$ and $T_\mathrm{B}$)
are easily seen to be anticorrelated, with the correlation coefficient
between them here being $-0.69$. Other strong correlations are clearly
apparent. For example, the block structure near the diagonal comes
about because the parameters within one group, such as the
normalization of the foreground temperatures in the different regions,
$T_0^i$, are strongly correlated with each other. When a parameter
outside this group is varied, the foregrounds in each region will all
have to change in a similar way to compensate, introducing a
correlation.

Some of the other features of the covariance matrix are
straightforward to understand.  For example, the temperatures of
turning points C and D are strongly anticorrelated with the foreground
temperature, $T_0^i$, for all $i$: an overall increase in the
brightness temperature of the 21-cm signal can be compensated for by a
decrease in the brightness of the foregrounds in every region of the
sky. This very strong anticorrelation may help to explain why the
inferred temperatures of the turning points become positively
correlated with each other, an effect which is evident in many of our
figures. It is more difficult to find `interesting' constraints on the
temperatures of the turning points than on the frequencies. The
similar temperature offsets of the different turning points for any
given noise realization may, however, allow us to recover the overall
shape of the signal well, even if its absolute normalization is
uncertain. The anticorrelation between the temperature of turning
point B and the foreground temperature is less strong than for the
other turning points, but this is mainly because of the larger
statistical error on the temperature of turning point B.

The difficulty of pinning down the overall normalization of the 21-cm
signal might be mitigated somewhat if we could fix its temperature at
some frequency using external or theoretical constraints. To some
extent we do this already by fixing the positions of turning points A
and E, which lie outside the observed band, and this appears to be
insufficient. The best candidate for a normalizing point inside the
{\it DARE} band is probably turning point D: looking at
Equation~\eqref{eqn:deltatb}, if $x_\mathrm{HI}\approx 1$
(reionization not yet seriously under way) and $T_\mathrm{S}\gg
T_\gamma$ (heating has saturated), the other terms can be computed
from well-constrained cosmological parameters and could be assumed to
be known. Interferometric experiments may be able to shed some light
on the value of $x_\mathrm{HI}$ and $T_\mathrm{S}$ and hence provide a
normalization indirectly. An EDGES-like experiment might also span
both the frequency of turning point D and high frequencies at which
$x_\mathrm{HI}\ll 1$ so the signal is known. It would face similar
problems to our reference experiment in constraining the large-scale
spectral shape, however, and so it is not clear it could provide a
much better temperature for turning point D.

Some features of the correlation matrix are more subtle: for example,
there is a striking anticorrelation between the normalization of the
solar spectrum in the different sky regions, $T_0^{i,\mathrm{Sun}}$,
and the running of the spectral index of the diffuse foregrounds,
$a_2^i$. This appears to come about because of the inverted spectrum
of the Sun relative to the spectrum of the diffuse foregrounds:
increasing $T_0^\mathrm{Sun}$ has a larger relative effect at high
frequency, where the diffuse foregrounds are weaker, and so the
spectrum of the diffuse foregrounds is made steeper at high
frequencies to compensate.

Including the effect of the Sun also impacts the correlation structure
of the other foreground parameters.  In this simulation, we assumed
that the contribution of the Sun to sky areas 1 and 2 (rows 63 and 64)
was identically zero, because these areas were observed while the Sun
was occluded by the Moon. This leads to the obvious stripe at this
position in the correlation matrix. One can easily see that the
correlations between the parameters of the diffuse foregrounds in
areas 1 and 2 are stronger than for the other sky areas: they have
less freedom to vary independently when there is no solar contribution
to take up the slack. This feature, and the anticorrelation between
$a_2^i$ and $T_0^{i,\mathrm{Sun}}$, justifies our assertion in
Section~\ref{subsec:otherfg} that it is important to include the
effect of the Sun in the modelling.

Degeneracies between the instrumental parameters other than
$\beta(\nu)$ (rows 41--53) appear to be very complex. This may be an
artefact of our parametrization of $\Gamma(\nu)$ in terms of DCT
coefficients, though it is hard to know in the absence of a more
physically motivated parametrization. Our main results assume tighter
priors on these parameters than were used to make
Fig.~\ref{fig:scaled_covmat}, which would make their correlations with
the foreground and signal parameters less important. Although beyond
the scope of this paper, it is possible that some alternative
instrument design would produce smaller degeneracies between
instrument and signal parameters, so that this sort of correlation
analysis might help in optimizing the instrument design. This could be
quite dependent on the signal model and parametrization though, and at
present it seems better to concentrate on producing a smooth
instrument response that can be described by a small number of
parameters.

\subsection{Comparison to other work}\label{subsec:comp}

In this paper, we have made use of an MCMC approach to estimate
constraints on the 21-cm global signal.  There has been a certain
amount of previous work making use of Fisher matrix approximations to
the likelihood, in the restricted case that the experiment genuinely
sees the full sky.  The initial work by \citet{sethi2005} in this area
assumed that foregrounds could be removed separately and completely
and so led to very optimistic predictions for cosmological
constraints. More in line with our approach here, \citet{PRI10}
accounted for the need to simultaneously fit the foregrounds and the
signal and introduced the turning point parametrization that we have
used throughout this paper.  Most recently, \citet{MOR11} investigated
constraints on more general models of reionization.

The confidence regions obtained using strong priors on the
instrumental and non-diffuse foreground parameters, and assuming 3000
h of data collection, are comparable to those found by \citet{PRI10}
for a 500 h observation of a single sky area (their figure
11). Fitting a model with many more parameters, as we do here, clearly
degrades the constraints we can obtain on the parameters of
interest for a given amount of integration time. Encouragingly,
though, this comparison shows that the degradation is not
catastrophic, and observing for a factor of a few longer allows us
to recover the loss.

It would be desirable to compare to the larger body of work
concentrating on probing the epoch of reionization with the global
21-cm signal \citep[e.g.][]{MOR11}, using appropriate models for the
frequency response (e.g.\ that of EDGES) and for the observational
strategy, which is somewhat different for ground-based
experiments. While our technique is applicable for models of the 21-cm
signal other than the turning point parametrization used here, the
signal during reionization is likely to be much more degenerate with
the foregrounds and instrumental response than the turning point
model. We defer a test of this statement to future work.

Probes of reionization other than the 21-cm line were studied by
\citet*{PLW10}, who discussed what current astrophysical priors can
tell us about reionization. Their framework could easily be extended
to account for global 21-cm experiments. Constraints from e.g.\ the
cosmic microwave background and the Ly$\alpha$ forest would not
necessarily be applicable directly to the positions of the turning
points in the parametrization we use here. Rather, constraints on the
turning points from global 21-cm experiments could be transformed into
constraints on the underlying physical model (the star formation
history, the efficiency of X-ray production, etc.), and the other
astrophysical constraints would also be applied in that space.

\section{Conclusions}\label{sec:conc}

We have presented a model for the data from a proposed lunar-orbiting
satellite to measure the global, redshifted 21-cm signal between 40
and 120 MHz. Fitting the parameters of this model to a realistic
simulated data set using an MCMC algorithm yields constraints on the
21-cm signal that are comparable to those found using much simpler
models for the foregrounds and instrument, despite the fact that we
use the data to constrain $\approx\! 73$ parameters, rather than 10.
The key assumptions used in extracting the signal are that the
foregrounds are smooth, that the instrumental response is also smooth
and can be determined reasonably well by independent measurements, and
that the 21-cm signal, averaged over the solid angle of our antenna
beam, is constant across the sky while the foregrounds are not.

A mission of reasonable duration ($\sim 3$ yr) can find the position
of the bottom of the `cosmic dawn' absorption trough in our fiducial
model with an accuracy of around $\pm 1\ \mathrm{MHz}$ in frequency
and $\pm 20\ \mathrm{mK}$ in temperature (2-$\sigma$ errors), provided
that the instrumental response has been well characterized. The
frequency position of the peak in emission at the onset of
reionization can be found to within $\pm 0.5\ \mathrm{MHz}$, while
`turning point B', marking the onset of Ly$\alpha$ pumping, can be
determined with a 1-$\sigma$ error of around $\pm 2.5\
\mathrm{MHz}$. For a shorter mission, of e.g.\ 1000 h, these
constraints degrade somewhat, and it may only be possible to find an
upper limit on the frequency of turning point B. A mission of 10000 h
allows a good measurement of the frequency of turning point B, with a
2-$\sigma$ confidence interval that lies entirely within the {\it
DARE} frequency band.

We have examined the effect of using prior information on the
non-diffuse foregrounds, which may be amenable to measurement from the
ground, and on the instrumental parameters. Priors on the foregrounds
do not help a great deal, though clearly it will still be valuable to
have independent, ground-based measurements of the foregrounds, to
inform our modelling and to check that our measurements are
consistent.  Tightened priors on the instrument model, however, which
correspond to improved calibration, reduce the statistical errors.
They may also help to reduce the importance of temperature errors
which are correlated across the frequency band, and which result in an
uncertainty in the overall normalization of the 21-cm brightness
temperature. Even if this correlation is present, it is likely that
the shape of the 21-cm signal can be recovered accurately, since the
absolute error on the temperature of each of the three turning points
tends to be similar.

Interferometric experiments must also perform foreground subtraction,
to an accuracy of around one part in $10^3$ for the diffuse Galactic
emission, and to one part in $10^6$ or even $10^8$ for bright point
sources \citep*{datta2010}. In this sense, arrays such as MWA, the Low
Frequency Array (LOFAR) and the Precision Array to Probe the Epoch of
Reionization (PAPER) should characterize the properties of the
Galactic and extragalactic foregrounds and validate the assumption
that they can be modelled using functions which deviate little from
power laws. They differ from experiments such as {\it DARE} or EDGES,
however, in that they will use observations of specific point sources
for calibration of gain and bandpass in a way that is not possible for
sky-averaged experiments, since the latter cannot isolate the
contribution to the measured spectrum from an individual source. For
this reason, upcoming arrays have not been designed to achieve an
intrinsically smooth bandpass that can be quantified with only a few
parameters, and therefore will likely shed little light on the
calibration or instrument modelling for sky-averaged experiments.

In this paper, we have focused on a particular reference experiment to
illustrate our techniques.  The methodology developed here is very
general and can easily be extended to other global 21-cm experiments.
As global 21-cm experiments continue to improve from their current
relative infancy, there will be a need for improved techniques of
statistical analysis. We have taken some early steps in that
direction.

It is worth reiterating, however, that there are several other stages
in the data analysis which must be passed before the methodology of
this paper can be applied. Individual spectra taken with a short
cadence (of e.g.\ 1~s) must be combined together using a map-making
procedure to produce something like the eight independent spectra seen
here. The frequency response must be internally calibrated, for
example by toggling the receiver input between the antenna feeds and
calibration loads. Narrow features such as RRLs or, in the case of
ground-based experiments, RFI, must be excised. All these steps become
more complicated for ground-based experiments. For the map-making, an
experiment fixed to the ground would not have the complete control
over the pointing direction provided by a satellite, and would not
have access to the whole sky. Moreover, the ionosphere effectively
causes the sky seen by the antenna to vary with time. Internal
calibration is made more awkward by changes in temperature and
atmospheric conditions, while a space environment is more predictable.
Finally, RFI is likely to be considerably more prevalent than RRLs.
The main effects of these earlier steps on the MCMC method are likely
to be the introduction of non-Gaussianity to the noise on the
frequency spectra, and correlation between different sky areas, both
of which affect the computation of the likelihood. It is not clear
whether some of these effects could be captured with extra nuisance
parameters in the MCMC. It will be important to study the preliminary
analysis steps and their impact on the final extraction step in future
work, especially if our formalism is to be adapted for use with
ground-based experiments.

\section*{Acknowledgments}

We thank Stuart Bale for providing an estimate of the noise caused by
exospheric dust impacts on the antenna and the spacecraft. We also
acknowledge the work of the {\it DARE} team in designing the mission,
including Joseph Lazio, Rich Bradley, Chris Carilli, Steve Furlanetto,
Avi Loeb, Larry Webster, Jill Bauman and Ian O'Dwyer. The authors are
members of the LUNAR consortium (http://lunar.colorado.edu),
headquartered at the University of Colorado, which is funded by the
NASA Lunar Science Institute (via Cooperative Agreement NNA09DB30A) to
investigate concepts for astrophysical observatories on the Moon.

\bibliography{darebib}

\end{document}